\documentclass[12pt]{iopart}
\usepackage{graphics,amssymb}

\begin{document}

\title[Netons: Vibrations of Complex Networks]
{Netons: Vibrations of Complex Networks}

\author{Beom Jun Kim\dag, H Hong\ddag, M Y Choi\P\ddag}
\address{\dag Department of Molecular Science and Technology, Ajou University,
Suwon 442-749, Korea}
\address{\ddag Korea Institute for Advanced Study, Seoul 130-012, Korea}
\address{\P Department of Physics, Seoul National University, Seoul 151-747, Korea}

\ead{beomjun@ajou.ac.kr}

\begin{abstract}
We consider atoms interacting each other through the topological structure of a
complex network and investigate lattice vibrations of the system,  the
quanta of which  we call {\em netons} for convenience.
The density of neton levels, obtained numerically, reveals that
unlike a local regular lattice, the system develops a gap of a finite width, 
manifesting extreme rigidity of the network structure at low energies.  
Two different network models, the
small-world network and the scale-free network, are compared:
The characteristic structure of the
former is described by an additional peak in the level density whereas a
power-law tail is observed in the latter, indicating excitability of netons at
arbitrarily high energies. The gap width is also found to vanish in the
small-world network when the connection range $r = 1$.
\end{abstract}


\maketitle

A variety of network systems such as computer networks, neuronal networks,
biochemical networks and social networks, possesses complex topological
structure, which can be described neither by regular networks nor by completely
random networks~\cite{ref:review}.  Among various models for generating such
complex networks, the Watts-Strogatz (WS) model~\cite{ref:WS} and the
Barab{\'a}si-Albert (BA) model~\cite{ref:BA} provide the two most
representative ones.  Both model networks are characterized by very small
network diameters proportional to the logarithm of the network size, which is
dubbed ``small-world behavior'' and commonly observed in various real networks.
Other important properties of many real network systems include high
clustering~\cite{ref:WS} and scale-free degree distributions~\cite{ref:BA}.
While most existing studies of complex networks have focused on the geometrical
and topological characterization, there now emerges an increasing number of
studies paying attention to the dynamics defined on the networks. 

Lattice vibrations and associated phonon excitations on local regular networks, e.g.,
one-dimensional (1D) chains or two-dimensional square lattices, have been a
textbook example of the introductory solid-state physics.  In the case of such
regular networks, atoms located on vertices interact only with nearest neighboring
atoms, resulting in the standard linear phonon dispersion.  On the other hand,
when the interaction between atoms takes the form of the connection topology of
a complex network, ingredients of long-range nature enter into the system.  For
example, the WS network has both local edges and long-range
shortcuts~\cite{ref:WS}, which in this work are interpreted as the coexistence
of the local and the long-range interactions.  
Existing studies on phonon
excitations have mostly been performed on the local regular networks such as
$d$-dimensional hypercubic lattices or on the fractal geometry~\cite{fracton}.
Recently, the spectral properties of lattice vibrations have been studied in 
a small-world network similar to the WS network~\cite{monasson}. However,
effects of the local interaction range and the dependence of the spectral
properties on the rewiring probability have not been investigated in detail.

The main purpose of this work is to study in detail the vibration spectra of 
both the WS network and the BA network, in comparison with those of regular lattices. 
We thus consider atoms located on vertices of a complex network and
interacting with each other connected via edges of the network, and investigate
lattice vibrations and phonon excitations, with emphasis on the effects of the
network topology.  The results of such study can have implications in
the mechanical property of a system of long chains bent in a complicated way.  
For example, in a bundle of long flexible polymer chains, 
some monomers which were separated by a long distance along a chain 
can make new couplings, building shortcuts. 
Throughout this work, we refer to phonons in a complex
network as {\em netons}, to manifest qualitatively different characteristics
due to the fact that the underlying geometric structure is not a usual periodic lattice.
By means of numerical diagonalization of the dynamic matrix,
neton excitation spectra are obtained and the corresponding density of neton
levels are computed.  It is revealed that the complex network in general has a
finite gap in the level density, indicating the absence of low-energy
excitations.  The dependence of the gap width on the rewiring probability
and the local interaction range is investigated in detail.
The existence of the gap is also reflected by the vanishingly
small specific heat at low temperatures.  In the WS network three singularities
are observed in the neton level density: Among them two are shown to be
reminiscence of the van Hove singularity present in the local regular network,
while the third one originates from the complex network structure.  The BA
network is also considered and the scale-freeness results in a power-law
distribution of the level density, suggesting excitability of netons with
arbitrarily high energies.

We build the WS network following Ref.~\cite{ref:WS}: First, a 1D regular
network with only local connections of range $r$ is constructed under the
periodic boundary conditions.  Next, each local edge (or link) is visited once,
and with the rewiring probability $P$ removed and reconnected to a randomly
chosen vertex.  After a whole sweep over the entire network, the
average number of shortcuts in the system of size $N$ is given by $NPr$.  In
the WS network built as above, an atom is put on every vertex whereas an edge
connecting two vertices represents the coupling between the two atoms located
at the two vertices.  For simplicity, we assume that all atoms are identical,
each having mass $M$ and moving only along the direction of the chain.  The
equation of motion for the $l$th atom at the position $x_l$ then reads
\begin{equation}
M\ddot x_l = C\sum_{m\in \Lambda_l}(x_m - x_l),
\label{eq:motion}
\end{equation}
where $\Lambda_l$ stands for the set of vertices connected to
the $l$th vertex (via either local edges or shortcuts) and
the interaction strength $C$ is assumed to be the same for any
pairs of the interacting atoms.

In the absence of shortcuts, Eq.~(\ref{eq:motion}) describes a 1D chain of
atoms with only local connections up to the $r$th nearest neighbors.  The
analytic solution in this limit of $P=0$ is easily found to take the form:
$ x_l \propto e^{i(kla-\omega t)}$, 
where $a$ represents the interparticle spacing, $\omega$ is the eigenfrequency
and the periodic boundary conditions, $x_{N+l}=x_l$, restrict the wavenumber
$k$ to take the value $k=2\pi n/N$ with integer $n$.  For simplicity, we set
$a\equiv 1$, and obtain 
\begin{equation}
\omega^2 = \frac{2C}{M}\sum_{j=1}^{r}(1-\cos j k),
\label{eq:w2}
\end{equation}
which leads to the dispersion relation
\begin{equation}
\omega \approx  \sqrt{\frac{Cr(r+1)(2r+1)}{6M}} k 
\label{eq:rsmall}
\end{equation}
for small wave numbers ($k \rightarrow 0)$.  Equation~(\ref{eq:rsmall}) shows
that $\omega \propto k$ regardless of the local interaction range $r$.  Such
linear dependence of the frequency $\omega$ on the wavenumber $k$ implies that
the group velocity equals the phase velocity, both of which are frequency
independent.  Henceforth we measure the frequency $\omega$ in units of
$\sqrt{C/M}$ for convenience.

In the presence of shortcuts $(P\neq 0)$, 
the plane wave solution fails, 
owing to the lack of the translational symmetry. 
Substituting $x_l = {\bar x}_l e^{i\omega t}$ in Eq.~(\ref{eq:motion}), 
we write the equation of motion in the matrix form 
\begin{equation}
\omega^2 \mathbf{X} = \mathbf{D X},
\label{eq:matrix}
\end{equation}
where $\mathbf{X}$ is the $N$-dimensional column vector with 
the components ${\bar x}_i$ $(i=1, \cdots, N)$.
The $N\times N$ dynamic matrix $\mathbf{D}$ has the elements   
\begin{equation} \label{eq:Dij}
D_{ij} = \left\{
\begin{array}{cl}
|\Lambda_i|  & \mbox{for $i=j$,} \\
-1  & \mbox{for $j\in \Lambda_i$,} \\
0 & \mbox{otherwise,}
\end{array}
\right.
\end{equation}
where $|\Lambda_i|$ denotes the number of vertices connected directly to the
$i$th vertex via local edges or shortcuts.  It should be noted that $D_{ij}$ lacks the translational symmetry due to the existence of shortcuts. 
We have performed numerical
diagonalization of the dynamic matrix $\mathbf{D}$ for various network sizes
$N=100$, 200, 400, and 800, only to find insignificant difference between the
two largest sizes.  The eigenfrequencies found numerically for given network
realization are sorted in the ascending order and an integer label $n$ is
attached from the lowest eigenfrequency ($n=0$ for the smallest $\omega$, $n=1$
for the next smallest $\omega$, and so on).  To obtain better statistics, we
have also considered 1000-10000 different network realizations, over which
averages are taken. 

\begin{figure}
\centering{\resizebox*{!}{6cm}{\includegraphics{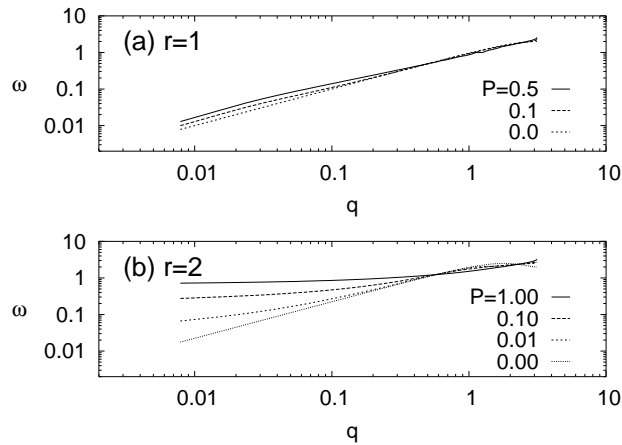}}}
\caption{Frequency spectra of the WS network with the size $N=400$ and the
local interaction range (a) $r=1$; (b) $r=2$.  The neton frequency $\omega$ is
plotted as a function of the quasi-wavenumber $q \equiv 2\pi n/N$, where $n$
($\geq 0$) is the integer labelling the eigenfrequency from the smallest one.
In the absence of shortcuts ($P=0$), the linear relation $\omega \sim q$ is
shown for both $r=1$ and $r=2$. In the presence of shortcuts ($P\neq 0$),
disclosed are very distinct frequency spectra between the two cases $r=1$ and
$r=2$.
}
\label{fig:wk}
\end{figure}

Figure~\ref{fig:wk} shows the neton frequency $\omega$ versus the {\em
quasi-wavenumber} $q\equiv 2\pi n/N$ at various values of the rewiring
probability $P$ for the local interaction range (a) $r = 1$ and (b) $r=2$.
Note that the quasi-wavenumber $q$ becomes identical to the true wavenumber $k$
only when $P=0$. In general, the wavenumber $k$ is not appropriate for the
system in the presence of shortcuts ($P \neq 0$): It does not make a good
quantum number for $P \neq 0$, where the translational symmetry is broken.
For the range $r=1$, Fig.~\ref{fig:wk}(a) suggests that the linear dispersion
relation $\omega \sim q$, which has been obtained analytically in
Eq.~(\ref{eq:rsmall}) for $P=0$, still appears to hold at any value of $P$.  In
contrast, for the local interaction range $r=2$, the dispersion $\omega(q)$
changes its functional form as $P$ is raised from zero, reducing to the linear
relation $\omega \sim q$ only in the limit of $P \rightarrow 0$ [see
Fig.~\ref{fig:wk}(b)].  In particular, it is evident in Fig.~\ref{fig:wk}(b)
that for $P\neq 0$ there appears a nonzero minimum value of $\omega$, below
which the eigenfrequency does not exist.  We have also examined the cases with
the range larger than two ($r>2$) and confirmed that these features remain
qualitatively the same.  With regard to this, it is of much interest to note
that the WS network has a finite clustering coefficient~\cite{footnote},
only when $r \geq 2$~\cite{ref:WS}, which makes it plausible that the difference in the neton
frequency spectrum between the cases $r=1$ and $r=2$, shown in
Fig.~\ref{fig:wk}(a) and (b), respectively, is a manifestation of the
clustering effects. 
We have also considered numerically the decomposition property of the network and 
found that while the WS network with $r \geq 2$ does not decompose into distinct parts, 
for $r=1$ the probability of decomposition is finite. 
However, it should be noted that those networks with $r=1$ which are not decomposed 
still display the vanishing gap.
Consequently, the vanishing gap in Fig.~\ref{fig:wk}(a) for $r=1$ has its 
origin not in the decomposition property but presumably in the clustering property 
of the network. 
Similar WS networks can also be built in a slightly
different manner: Instead of rewiring local connections with the total number
of edges fixed, one may {\it add} shortcuts to the network without removing
local connections.  In this case, we have confirmed numerically that there
exists a gap even for $r=1$, which keeps parallel to the observation that the
$XY$ model on the WS network with $r=1$, constructed in such a way,
has a finite-temperature transition, similar to that in the original 
WS network with $r>1$~\cite{ref:Hongnn}. 

We next consider the neton-level density $g(\omega)$, which is given 
by~\cite{ref:Solidstate} 
\begin{equation}
g(\omega) \propto \int dk~\delta\bigr(\omega-\omega (k)\bigl)
\label{eq:density1d}
\end{equation}
in one dimension. 
Using the mathematical identity 
$\delta\bigr(\omega-\omega(k)\bigl)= {|d\omega/dk|}^{-1} \delta(k-k_0)$
with $\omega(k_0)\equiv \omega$, we write Eq.~(\ref{eq:density1d}) in the form
\begin{equation}
\label{eq:gw}
g(\omega) \propto \left|\frac{d\omega}{dk}\right|^{-1}_{k_{0}}.
\end{equation}
In the regular network ($P=0$), Eq.~(\ref{eq:gw}), together with
Eq.~(\ref{eq:rsmall}), gives the density $g(\omega)$ independent of $\omega$,
i.e., $g(\omega) \sim \omega^0$, in the low-frequency limit $\omega \rightarrow
0$.

\begin{figure}
\centering{\resizebox*{!}{6.0cm}{\includegraphics{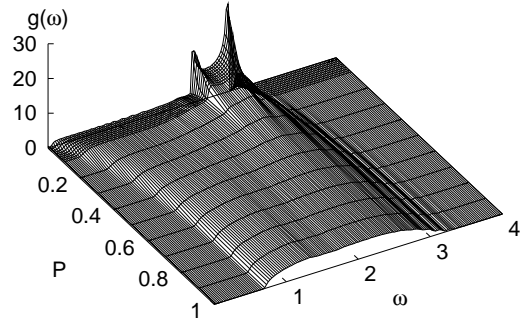}}}
\vskip -0.8cm
\caption{Neton-level density $g(\omega)$ of the WS network with
size $N=400$ and the local interaction range $r=2$ as a function of the frequency $\omega$ and
the rewiring probability $P$.
}
\label{fig:g(w)}
\end{figure}

\begin{figure}
\centering{\resizebox*{!}{6.0cm}{\includegraphics{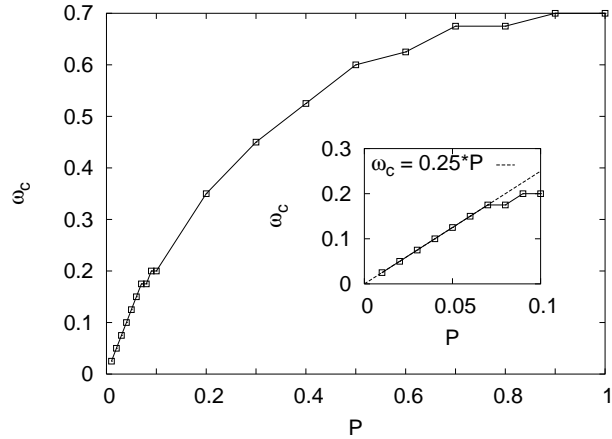}}}
\caption{Dependence of the gap width $\omega_c$ on the rewiring probability
$P$.  The inset demonstrates that $\omega_c$ is well described by the relation 
$\omega_c \propto P$ for small $P$. 
}
\label{fig:wc}
\end{figure}

\begin{figure}
\centering{\resizebox*{!}{6.0cm}{\includegraphics{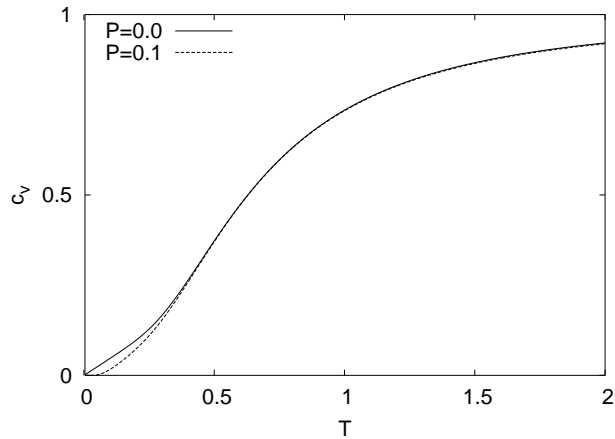}}}
\caption{Specific heat $c_v$ of the WS network with $N=400$ and $r=2$
at $P=0$ and $0.1$.  The existence of a gap in the neton-level density
at $P=0.1$ is reflected by the exponentially small specific heat at low 
temperatures.
}
\label{fig:WScv}
\end{figure}

In the presence of shortcuts, we use the box counting method to compute the
neton-level density from the obtained eigenvalues of the dynamic matrix.  The
resulting density $g(\omega)$ is plotted in Fig.~\ref{fig:g(w)}, which exhibits
several characteristic features.  First, as already found in
Fig.~\ref{fig:wk}(b), the density $g(\omega)$ develops a gap $\omega_c$ below
which $g(\omega) = 0$.  Figure~\ref{fig:wc} shows that the gap $\omega_c$, well
described by the linear relation $\omega_c \propto P$ for small $P$ (see the
inset of Fig.~\ref{fig:wc}), keeps increasing as the rewiring probability $P$
is raised and then appears to saturate for $P \gtrsim 0.7$.  Such saturation
behavior has also been observed in the synchronization phenomena in the WS
network~\cite{ref:Hong}.  The emergence of the gap implies that the WS network
is very rigid against lattice vibrations at small energies.  In other words,
one cannot excite netons at sufficiently low energies.  We also compute the
specific heat of the WS network for $P=0.1$, and compare it with that of the
local regular network ($P=0$) in Fig.~\ref{fig:WScv}.  As expected, the
presence of a finite gap in the neton-level density gives rise to
exponentially small specific heat at low temperatures.  This is in contrast to
the local regular network, where thermal excitations of phonons are gapless and
accordingly, the specific heat is described by the well-known Debye form with
power-law dependence on the temperature, i.e., $c_v \propto T$ in one
dimension.

\begin{figure}

\centering{\resizebox*{!}{6.0cm}{\includegraphics{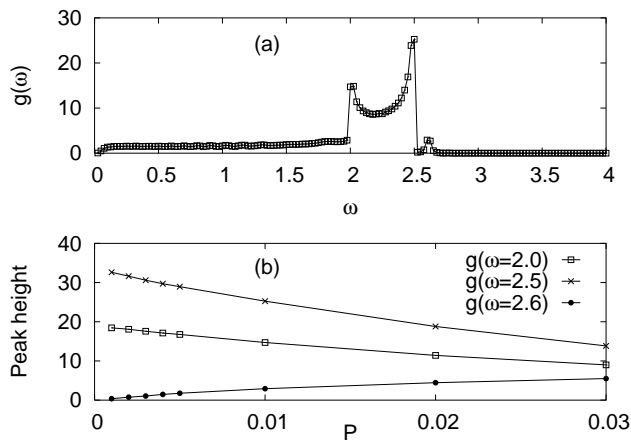}}}
\caption{(a) Neton-level density $g(\omega)$ in the WS network with size
$N=400$ for $r=2$ and $P=0.01$.  (b) Heights of the three peaks in (a) versus
$P$.  Among the three, two peaks are the reminiscence of the local regular
network and consequently their heights decrease with $P$.  In contrast, the
height of the third peak increases with $P$, suggesting that it has its origin
in the complex network structure.
}
\label{fig:vanHove}
\end{figure}

The second interesting feature observed in the neton-level density in
Fig.~\ref{fig:g(w)} is the emergence of sharp peaks at small $P$.
Figure~\ref{fig:vanHove}(a) shows in detail the density $g(\omega)$ at
$P=0.01$, displaying three peaks.  It is well known that a local regular
network ($P=0$) (i.e., a regular lattice) has van Hove singularities whenever
$d\omega/dk$ vanishes [see Eq.~(\ref{eq:gw})].  For example, in the local
regular network with the second-nearest neighbor interaction $(r=2)$, the
positions of the van Hove singularities are easily obtained: The eigenfrequency
$\omega=\sqrt{2(2-\cos k-\cos 2k)}$ given by Eq.~(\ref{eq:w2}), combined with
the condition $d\omega/dk = 0$, leads to two singularities at $\omega=2$ and
$\omega= 5/2 $; this allows us to conclude that the first two peaks in
Fig.~\ref{fig:vanHove}(a) are remnants of the van Hove singularities present in
the local regular network.  On the other hand, the third peak at $\omega
\approx 2.6$ cannot be explained in terms of the singularity of the local
regular network.  In order to probe the origins of the peaks, we measure how
the peak heights change as the rewiring probability $P$ is varied and plot the
results in Fig.~\ref{fig:vanHove}(b).  The first peak and the second one at
$\omega = 2$ and at $\omega = 2.5$, respectively, are observed to have heights
which decrease with $P$.  This again confirms that these two peaks originate
from the local regular network, thus becoming weaker as the network loses its
local structure.  The height of the third peak, in contrast, is found to grow
as $P$ is increased, which clearly indicates that the third peak provides a
feature generic to the WS network structure.

\begin{figure}
\centering{\resizebox*{!}{6.0cm}{\includegraphics{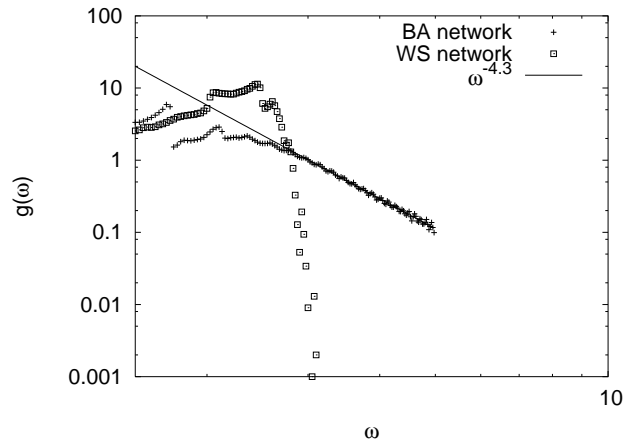}}}
\caption{Neton-level density $g(\omega)$ in the BA network with size $N=200$
and $M=3$ is compared with that in the WS network with $N=400$, $r=2$, and
$P=0.1$.  The log-log plot of $g(\omega)$ versus $\omega$ discloses the
power-law tail at high frequencies in the BA network.
}
\label{fig:BAtail}
\end{figure}

Finally, we examine neton excitations on the BA network~\cite{ref:BA}.  In
more detail, we first grow the BA network, following Ref.~\cite{ref:BA}:
Starting from $M$ vertices initially (at time $t=0$), we attach a vertex with
$M$ edges at every time step. Once the BA network is built, we then put an atom
on each vertex, assuming that the network is embedded in a 1D chain.  The
frequency spectrum is obtained in the same way as that of the WS network, with
the dynamic matrix still given by Eqs.~(\ref{eq:matrix}) and (\ref{eq:Dij}).
Similarly to the WS network, revealed is the emergence of a gap.  As the number
$M$ becomes larger, the gap grows and the peaks tend to smear out.  Another
interesting observation in the BA network is the slow decay of the density
$g(\omega)$ at large frequencies, which is to be compared with the vanishingly
small density at high frequencies found in the WS network.  The log-log plot of
the density $g(\omega)$ versus the frequency $\omega$, displayed in
Fig.~\ref{fig:BAtail}, reveals that the BA network exhibits power-law decay
which is well scaled as $g(\omega)\sim\omega^{-4.3}$.  In contrast, $g(\omega)$
in the WS network shows very rapid decay, which is presumably of the
exponential form.  The observed power-law decay of $g(\omega)$ in the BA
network implies that netons may be excited with arbitrarily high energies.  It
is of interest to compare this with the behavior of the adjacency matrix of the
BA network~\cite{ref:goh}.  Although the adjacency matrix $\mathbf{A}$, with
$A_{ii} = 0$ and $A_{ij} = -D_{ij}$ for $i \neq j$, and the dynamic matrix
$\mathbf{D}$ have different diagonal components, both show similar scale-free
distributions at large eigenvalues.

In summary, we have investigated the phonon spectra in the monatomic chain
where atoms interact with each other through the edges of complex networks.
The density of phonon levels in the complex network, i.e., neton levels
has been found to develop a gap, which indicates
that netons may not be excited up to the critical frequency, and accordingly,
a finite amount of energy is required to generate lattice vibrations.  This
robustness of a complex network has also been shown to yield exponentially
small specific heat at low temperatures, in contrast to the power-law
temperature dependence in the local regular network.  The characteristic
structure of the WS network has been described by the emergence of a peak in
the level density in addition to those originating from the van Hove
singularities.  In the BA network, on the other hand, the level density has
been found to decay algebraically at high frequencies, which suggests
excitability of arbitrarily high-energy netons.

\ack

This work was supported in part by the Korea Science and Engineering Foundation 
through Grant No.\ R14-2002-062-01000-0 (BJK) and 
in part by the Ministry of Education of Korea through the BK21 project (MYC).


\section*{References}

\begin{thebibliography}{99}

\bibitem{ref:review}
For reviews, see, e.g., the special issue on complex systems in 1999 {\it Science} {\bf 284} 79;
Watts D J, {\it Small Worlds} (Princeton Univ. Press, Princeton, 1999);
Newman M E J 2000 {\it J. Stat. Phys.} {\bf 101} 819;
Strogatz S H 2001 {\it Nature (London)} {\bf 410} 268;
Dorogovtsev S N and Mendes J  F F, {\it Adv. Phys.} {\bf 51}, 1079 (2002);
Albert R and Barab{\'a}si A-L 2002 {\it Rev. Mod. Phys.} {\bf 74} 47.

\bibitem{ref:WS}
Watts D J and Strogatz S H 1998 {\it Nature (London)} {\bf 393}, 440.

\bibitem{ref:BA}
Barab{\'a}si A-L and Albert R 1999 {\it Science} {\bf 286}  509;
Barab{\'a}si A-L, Albert R and Jeong H 1999 {\it Physica A} {\bf 272}  173;
Albert R, Jeong H and Barab{\'a}si A-L 1999 {\it Nature (London)} {\bf 401} 130.

\bibitem{fracton}
Alexander S and  Orbach R L 1982 {\it J. de Phys. Lett.} {\bf 43} L625.

\bibitem{monasson} Monasson R 1999 {\it Eur. Phys. J B} {\bf 12} 555.

\bibitem{footnote}
The clustering coefficient is defined to measure the extent to which 
two neighboring vertices of a vertex are also neighbors 
of each other (see Ref.~\cite{ref:WS}). 

\bibitem{ref:Hongnn}
Hong H,  Choi M Y and Kim B J 2002 {\it Phys. Rev. E} {\bf 65} 047104.

\bibitem{ref:Solidstate}
See, e.g., Ashcroft N W and Mermin I {\it Solid State Physics} (Saunders, N.Y., 1976).

\bibitem{ref:Hong}
Hong H,  Choi M Y and Kim B J 2002 {\it Phys. Rev. E} {\bf 65} 026139.


\bibitem{ref:goh}
Goh K-I, Kahng B and Kim D 2001 {\it Phys. Rev. E} {\bf 64} 051903.

\end{thebibliography}
\end{document}